\documentclass[superscriptaddress,nofootinbib,onecolumn,,longbibliography]{revtex4}
\usepackage{amssymb,amsmath}
\usepackage{mathtools}
\usepackage{hyperref}
\hypersetup{colorlinks=true, linkcolor=blue, citecolor=blue}

\DeclarePairedDelimiter\floor{\lfloor}{\rfloor} 

\begin{document}
	
	\title{Dimensional reduction of a finite-size scalar field model at finite temperature}
	\author{E. Cavalcanti}
	\email[]{erich@cbpf.br}
	\affiliation{Centro Brasileiro de Pesquisas F\'{\i}sicas/MCTI, Rio de Janeiro, RJ, Brazil}
	\author{J. A. Louren\c{c}o}
	\email[]{jose.lourenco@ufes.br}
	\affiliation{Universidade Federal do Esp\'{i}rito Santo, Campus S\~ao Mateus, ES, Brazil}
	\author{C.A. Linhares}
	\email[]{linharescesar@gmail.com}
	\affiliation{Instituto de F\'{\i}sica, Universidade do Estado do Rio de Janeiro, Rio de Janeiro, RJ, Brazil}
	\author{A. P. C. Malbouisson}
	\email[]{adolfo@cbpf.br}
	\affiliation{Centro Brasileiro de Pesquisas F\'{\i}sicas/MCTI, Rio de Janeiro, RJ, Brazil}
	
	\begin{abstract}
		We investigate the process of dimensional reduction of one spatial dimension in a thermal scalar field model defined in $D$ dimensions (inverse temperature and $D-1$ spatial dimensions). We obtain that a thermal model in $D$ dimensions with one of the spatial dimensions having a finite size $L$ is related to the finite temperature model with just $D-1$ spatial dimensions and no finite size. Our results are obtained for one-loop calculations and for any dimension $D$. For example, in $D=4$ we have a relationship between a thin film with thickness $L$ at finite temperature and a surface at finite temperature. We show that, although a strict dimensional reduction is not allowed, it is possible to define a valid prescription for this procedure. 
	\end{abstract}
	
	\maketitle
	
	\section{Introduction}
	
	
	
	
	
	Dimensional reduction states that considering a model in $D$ dimensions we can obtain -- following some prescription -- a theory with fewer dimensions. One approach is to take a model with a compactified dimension and investigate its behavior when the size of the compactified dimension is reduced to zero.
	
	Let us consider the case of a finite-temperature field theory using the imaginary-time formalism. Temperature is introduced by compactifying the imaginary-time variable using periodic or antiperiodic boundary conditions, respectively, for bosons or fermions. In this context, dimensional reduction would be equivalent to a high-temperature limit. This idea was proposed long ago by Appelquist and Pisarski~\cite{Appelquist:1981vg} and later formulated in more detail by Landsman~\cite{Landsman:1989be}. This idea is currently accepted and understood~\cite{zinnjustin2002,MeyerOrtmanns:2007qp}. However, up to now, we think that there is a lack of investigation on this topic when one is interested in a greater number of compactifications.
	
	We might ask how a system in arbitrary $D$ dimensions with two compactified dimensions, one corresponding to the inverse temperature $\beta = 1/T$ and the other with a finite-size $L$, behave when $L \rightarrow 0$. 
	However, strictly speaking, in the context of both first-order and second-order phase transitions, it can be shown that the limit of $L\rightarrow 0$ cannot be fully attained~\cite{Khanna:2014qqa}. There are many works in the context of phase transitions in thin films using different models (Ginzburg-Landau, Nambu-Jona-Lasinio, Gross-Neveu and others), see Refs.\cite{Linhares:2006my,Abreu:2003zz,Abreu:2009zz,Abreu:2011rj,Abreu:2011zzc,Linhares:2011nh,Linhares:2012vr,Khanna:2012zz,Khanna:2012js,Abreu:2013nca,Correa:2013mta}, that indicate the existence of a minimal thickness below which no phase transition occur. There is also, see for instance Ref.\cite{Linhares:2006my}, a comparison between a phenomenological model for superconducting thin films using a Ginzburg-Landau model and experimental results that strengthen the indication. Therefore, this is a direct indication that the physics of surfaces and thin films are different and one could not achieve a surface from a thin film.
	
	In this article, we investigate the problem of dimensional reduction from a mathematical physics perspective, for example, to investigate the relationship between systems in the form of ``films'' and ``surfaces''. Here this is generalized so that we investigate the relation between a $D$-dimensional and a $D-1$-dimensional scenario.
	
	To study a quantum field theory at finite temperature and finite size we use the formalism of quantum field theory in spaces with toroidal topologies~\cite{Khanna:2014qqa,Khanna:2009zz}. As a first investigation, we consider a scalar field model at the one-loop level. We obtain a remarkably simple relationship between the following situations: 
	\begin{itemize}
		\item Starting with a space in $D$ dimensions, we consider two compactifications: one of the dimensions corresponding to the inverse temperature $1/T$, the other one to a finite size $L$, and the other $D-2$ dimensions remaining of infinite size.
		\item Another possibility is to eliminate from the beginning one spatial dimension, and starting in a space with $D-1$ dimensions, we consider one compactification corresponding to the inverse temperature $1/T$.
	\end{itemize}
	This should, for example, entail a relation between surfaces and thin films for $D=4$. Furthermore, from a mathematical physics perspective, we also investigate the possibility of fractal dimensions. 
	
	\section{The model}
	
	In this article, we take a scalar field theory with a quartic interaction in $D$ dimensions with euclidean action
	\begin{equation}
	S_E = \int d^D x \left\{ \frac{1}{2}(\partial \phi)^2 + \frac{M^2}{2} \phi^2 + \frac{\lambda}{4!} \phi^4 \right\}. 
	\end{equation}
	\noindent We shall only discuss the 1-loop Feynman amplitude $\mathcal{I}$ with $\rho$ propagators and zero external momenta,
	\begin{equation}
	\mathcal{I}_\rho^D(M^2) = \int \frac{d^D p}{(2\pi)^D} \frac{1}{(p^2+M^2)^\rho}.
	\end{equation}
	\noindent In particular, $\rho=1$ corresponds to the tadpole contribution to the effective mass and $\rho=2$ corresponds to the first-order correction to the coupling constant.
	We introduce periodic boundary conditions on $d<D$ coordinates. The compactification of the imaginary time introduces the inverse temperature $\beta = 1/T \equiv L_0$ and the compactification of the spatial coordinates introduces the characteristic lengths $L_i$. So that the amplitude becomes
	\begin{equation}
	\mathcal{I}_\rho^{D,d}(M^2; L_\alpha)= \frac{1}{\prod_{\alpha=0}^{d-1} L_\alpha} \sum_{n_0,\ldots,n_{d-1}=-\infty}^{\infty} \int \frac{d^{D-d} q}{(2\pi)^{D-d}} \frac{1}{[q^2 + M^2 + \sum_{\alpha=0}^{d-1} (\frac{2\pi n_\alpha}{L_\alpha})^2]^\rho}.
	\end{equation}
	We compute the remaining integrals on the $(D-d)$-dimensional subspace using dimensional regularization. The remaining infinite sum can be identified as an Epstein-Hurwitz zeta function~\cite{Elizalde:1996zk} and leads -- after an analytic continuation -- to the sum over modified Bessel functions of the second kind $K_\nu(x)$; see Refs.~\cite{Cavalcanti:2017wnm,Khanna:2014qqa} for further details. The function $\mathcal{I}_\rho^{D,d}(M^2; L_\alpha)$ reads in the case of $d=2$
	\begin{equation}
	\mathcal{I}_\rho^{D,2}(M^2; \beta,L)
	= \frac{(M^2)^{-\rho+\frac{D}{2}} \Gamma\left[\rho-\frac{D}{2}\right]}{(4\pi)^{\frac{D}{2}}\Gamma[\rho]} + \frac{{\mathcal W}_{\frac{D}{2}-\rho} \left(M^2; \beta,L\right)}{(2\pi)^{\frac{D}{2}} 2^{\rho-2}\Gamma[\rho]},
	\label{Eq:FuncId2}
	\end{equation}
	\noindent where
	\begin{equation}
	{\mathcal W}_\nu (M^2;\beta,L) = \sum_{n=1}^\infty \left(\frac{M}{n\beta}\right)^\nu\hspace{-5pt}K_\nu(n\beta M)
	+\sum_{n=1}^\infty \left(\frac{M}{nL}\right)^\nu\hspace{-5pt} K_\nu(nL M)
	+2\hspace{-7pt}\sum_{n_0,n_1=1}^\infty \left(\frac{M}{\sqrt{n_0^2\beta^2+n_1^2L^2}}\right)^\nu\hspace{-5pt} K_\nu\left(M\sqrt{n_0^2\beta^2+n_1^2L^2}\right)
	\label{Eq_Ap:Wnu_def}
	\end{equation}
	with $\nu=\frac{D}{2}-\rho$. 
	
	As in Ref.~\cite{Fucci:2014mya}, we investigate these infinite sums using a representation of the modified Bessel function in the complex plane,
	\begin{equation}
	K_\nu(X) = \frac{1}{4\pi i} \int_{c-i\infty}^{c+i\infty} dt\; \Gamma(t)\Gamma(t-\nu)\left(\frac{X}{2}\right)^{-2t+\nu}.
	\label{Eq_Ap:Knu_complexplane}
	\end{equation}
	\noindent We remark that $c$ is a point located on the positive real axis that has a greater value than all those of the poles of the gamma function. It is clear in this definition that $c > \text{max}[0,\nu]$. However, we extend this definition so that we are allowed to interchange the integral over $t$ and the summation over $n$. Therefore, $c$ must be chosen in such a way that there is no pole located on its right. Substituing Eq.~\eqref{Eq_Ap:Knu_complexplane} in Eq.~\eqref{Eq_Ap:Wnu_def} we obtain
	\begin{multline}
	{\mathcal W}_\nu(M^2;\beta,L) = \frac{1}{4\pi i} \int_{c-i\infty}^{c+i\infty} dt\; \Gamma(t)\Gamma(t-\nu)\zeta(2t)\left(\frac{M^2}{2}\right)^{\nu} \left[\left(\frac{M\beta}{2}\right)^{-2t}+\left(\frac{ML}{2}\right)^{-2t}\right]
	\\+ \frac{1}{2\pi i} \int_{c-i\infty}^{c+i\infty} dt\; \Gamma(t)\Gamma(t-\nu)\left(\frac{M^2}{2}\right)^{\nu}\left(\frac{M}{2}\right)^{-2t} \sum_{n_0,n_1=1}^\infty \left(n_0^2\beta^2+n_1^2L^2\right)^{-t}.
	\label{Eq:Wnudef2}
	\end{multline}
	The double sum in Eq.~\eqref{Eq:Wnudef2} is known and has the analytical extension \cite{Elizalde:1996zk}
	\begin{equation}
	\sum_{n_0,n_1=1}^\infty \left(n_0^2\beta^2+n_1^2L^2\right)^{-t} = - \frac{\zeta(2t)}{2L^{2t}} + \frac{\sqrt{\pi}L}{2\beta} \frac{\Gamma(t-\frac{1}{2})\zeta(2t-1)}{\Gamma(t)L^{2t}}
	+ \frac{2\pi^t}{\Gamma(t)} \sum_{n_0,n_1=1}^\infty \left(\frac{n_0}{n_1}\right)^{t-\frac{1}{2}} \sqrt{\frac{L}{\beta}} \frac{1}{(\beta L)^t} K_{t-\frac{1}{2}} \left(2\pi n_0n_1 \frac{L}{\beta}\right).
	\label{Eq:DoubleSum}
	\end{equation}
	\noindent 
	After substituting Eq.~\eqref{Eq:DoubleSum} in Eq.~\eqref{Eq:Wnudef2} and also taking into account Eq.~\eqref{Eq_Ap:Knu_complexplane} we finally obtain the integral representation in the complex plane for the function $\mathcal{W}_\nu$ as the sum of these terms:
	\begin{align}
	\left(\frac{2}{M^2}\right)^{\nu} {\mathcal W}_\nu &= {\mathcal W}_\nu^{(1)} + {\mathcal W}_\nu^{(2)} + {\mathcal W}_\nu^{(3)}, \text{where}
	\label{Eq_Ap:Wnu_full}\\
	{\mathcal W}_\nu^{(1)} &= \frac{1}{4\pi i} \int_{c-i\infty}^{c+i\infty}\hspace{-10pt} dt\; \Gamma(t)\Gamma(t-\nu)\zeta(2t) \left(\frac{M\beta}{2}\right)^{-2t},
	\nonumber\\
	{\mathcal W}_\nu^{(2)} &= \frac{\sqrt{\pi}L}{4\pi i\beta}\int_{c-i\infty}^{c+i\infty}\hspace{-10pt} dt\; \Gamma(t-\nu)\Gamma\left(t-\frac{1}{2}\right)\zeta(2t-1)\left(\frac{ML}{2}\right)^{-2t},
	\nonumber\\
	{\mathcal W}_\nu^{(3)} &= \frac{1}{2\pi i} \sum_{k=0}^\infty \frac{(-1)^k}{\Gamma(k+1)} 
	\left(\frac{M \beta}{2\pi}\right)^{2k-2\nu}
	\int_{c-i\infty}^{c+i\infty}\hspace{-5pt} \frac{dt\;}{\sqrt{\pi}} \Gamma(t) \zeta(2t) 
	\Gamma\left(t-\nu+k+\frac{1}{2}\right) \zeta(2t-2\nu+2k+1) \left(\frac{\pi L}{\beta}\right)^{-2t}.
	\nonumber
	\end{align}
	
	The next step is to determine position of the poles of the functions in the complex-plane integrals and compute their residues. We see that the positions of the poles depend on the value of $\nu$. We compute the integrals for the following specific cases: 
	\begin{itemize}
		\item integer $\nu$, related to an even number of dimensions $D$; 
		\item half-integer $\nu$, related to an odd number of dimensions $D$; 
		\item and, for completeness, other real values of $\nu$, which can be thought of as related to fractal dimensions $D$.
	\end{itemize}
	
	We consider during this article two different scenarios. The first case considers a model in $D$ dimensions with two compactified dimensions, one related to the inverse temperature $\beta=1/T$ and the other with a finite size $L$. For this one we need the function ${\mathcal W}_\nu$, as defined in Eq.~\eqref{Eq_Ap:Wnu_full}, and obtain Eq.~\eqref{Eq:FuncId2}.
	\noindent On the other situation, we consider a model in $D-1$ dimensions with just one compactified dimension which is related to the inverse temperature. We see that, the amplitude requires just the knowledge of  ${\mathcal W}_\nu^{(1)}$,
	\begin{equation}
	\mathcal{I}^{D-1,1}_\rho(M^2; \beta)
	= \frac{\left(\frac{M^2}{2}\right)^{\frac{D}{2}-\rho}}{(2\pi)^{\frac{D}{2}} 2^{\rho-2}\Gamma[\rho]} \left\{
	\frac{\sqrt{\pi}}{2M}\Gamma\left(\rho-\frac{D}{2}+\frac{1}{2}\right)
	+ \frac{2\sqrt{\pi}}{M} {\mathcal W}^{(1)}_{\frac{D}{2}-\frac{1}{2}-\rho} \left(M^2; \beta\right)
	\right\}.
	\label{Eq:FuncId1}
	\end{equation}
	
	To avoid a lengthy exposition we exhibit for each case only the final expressions. For integer $\nu$ the function ${\mathcal W}_\nu^{(1)}$ reads
	\begin{multline}
	{\mathcal W}_\nu^{(1)} = \frac{\sqrt{\pi}}{2M\beta} \Gamma\left(\frac{1}{2}-\nu\right) 
	+ \frac{1}{2} \mathcal{S}^{(3)}_0 \left(\nu;\frac{M\beta}{2}\right)
	+ \frac{1}{2} \mathcal{S}^{(2)}\left(\nu;\frac{M\beta}{4\pi}\right)\\
	+ \begin{cases}
	- \frac{1}{4} \Gamma(-\nu) & \nu<0 \\
	- \frac{(-1)^\nu}{4\Gamma(\nu+1)^2} {\nu+1 \brack 2}
	+ \frac{(-1)^\nu}{2\Gamma(\nu+1)} \left(\gamma+\ln \frac{M \beta}{4\pi}\right)& \nu\ge 0
	\end{cases},
	\label{Eq:Wnu_Partial_Integer}
	\end{multline}
	\noindent and ${\mathcal W}_\nu$ reads
	\begin{multline}
	\left(\frac{2}{M^2}\right)^\nu \mathcal{W}_\nu  =
	+ \frac{1}{2} \mathcal{S}^{(2)} \left(\nu; \frac{ML}{4\pi}\right)
	+ \frac{1}{2} \mathcal{S}^{(3)}_1 \left(\nu; \frac{ML}{2}\right)
	- \frac{2\pi}{M^2 \beta L} \left\{
	\mathcal{S}^{(1)} \left(\nu; \frac{M\beta}{2\pi}\right)
	+ \mathcal{S}^{(4)} (\nu; M\beta)\right\}\\
	+\begin{cases}
	- \frac{1}{4} \Gamma(-\nu) & \nu<0\\
	+ \frac{(-1)^\nu}{2\Gamma(\nu+1)} \left\{
	- \frac{1}{2\Gamma(\nu+1)} {\nu+1 \brack 2}
	+ \gamma+\ln \frac{M L}{4\pi}
	+\frac{\pi}{6}\frac{\beta}{L}
	\right\} & \nu \ge 0\\
	+ \frac{\pi}{M^2\beta L} \Gamma(-\nu+1) & \nu<1\\
	+ \frac{\pi}{M^2\beta L} \frac{(-1)^\nu}{\Gamma(\nu)} \left\{
	- \frac{1}{\Gamma(\nu)} {\nu \brack 2}
	+  2 \ln M \beta
	-\frac{\pi}{3}\frac{\beta}{L}\right\} & \nu \ge 1
	\end{cases}.
	\label{Eq:Wnu_Full_Integer}
	\end{multline}
	\noindent Here the notation ${a\brack 2}$ indicates the unsigned Stirling number of the first kind: $${1\brack 2} = 0, {2\brack 2} = 1, {3\brack 2} = 3, {4\brack 2} = 11, {5\brack 2} = 50.$$ 
	\noindent In the above expressions, the functions $\mathcal{S}^{(1)}$,  $\mathcal{S}^{(2)}$, $\mathcal{S}^{(3)}$ and $\mathcal{S}^{(4)}$ are defined by
	\begin{align}
 	\mathcal{S}^{(1)}_i \left(\nu; \alpha\right) &= \sum_{k=1+i}^{\infty} (-1)^{k+\nu} \frac{\Gamma(k) \zeta(2k)}{\Gamma(k+\nu)} \alpha^{2k},\\
	\mathcal{S}^{(2)} \left(\nu; \alpha\right) &= \sum_{k=1}^{\infty} (-1)^{k+\nu} \frac{ \Gamma(2k+1) \zeta(2k+1)}{\Gamma(k+1) \Gamma(k+\nu+1)} \alpha^{2k},\\
	\mathcal{S}^{(3)}_i \left(\nu; \alpha\right) &= \sum_{k=1+i}^{\nu} (-1)^{\nu-k} \frac{ \Gamma(k)\zeta(2k)}{\Gamma(\nu-k+1)}\alpha^{-2k},\\
	\mathcal{S}^{(4)} \left(\nu; \alpha\right) &= \sum_{k=1}^{\nu-1} (-1)^{\nu-k} \frac{ \Gamma(2k+1)\zeta(2k+1)}{\Gamma(k+1)\Gamma(\nu-k)}\alpha^{-2k}.
	\end{align}
	
	For half-integer values of $\nu$ ($\nu = \mu + 1/2$, for integer $\mu$) we obtain
	\begin{multline}
	{\mathcal W}_\nu^{(1)} = 
	- \frac{1}{4} \Gamma\left(-\frac{1}{2}-\mu\right) 
	- \frac{\sqrt{\pi}}{M \beta} \mathcal{S}^{(4)} (\mu+1;M\beta) 
	- \frac{\sqrt{\pi}}{M \beta} \mathcal{S}^{(1)}_0 \left(\mu+1;\frac{M \beta}{2\pi}\right)\\
	+ \begin{cases}
	+\frac{\sqrt{\pi}}{2M \beta} \Gamma(-\mu) & \mu<0\\
	+ \frac{\sqrt{\pi}}{2M\beta} \frac{(-1)^\mu}{\Gamma(\mu+1)} \left( {1+\mu \brack 2} \frac{1}{\Gamma(\mu+1)} - \ln M \beta \right) &\mu \ge 0
	\end{cases},
	\label{Eq:Wnu_Partial_HalfInteger}
	\end{multline}
	and
	\begin{multline}
	\mathcal{W}_\nu \left(\frac{2}{M^2}\right)^\nu =  
	-\frac{1}{4} \Gamma\left(-\frac{1}{2}-\mu\right) 
	+\frac{\pi}{M^2 \beta L} \Gamma \left(\frac{1}{2}-\mu\right)
	+ \frac{\sqrt{\pi}}{ML} \frac{(-1)^\mu}{\Gamma(\mu+1)} \left(\gamma + \ln \frac{\beta}{4\pi L}\right) \\
	- \frac{\sqrt{\pi}}{M L} \mathcal{S}^{(1)}_0 \left(\mu+1;\frac{M L}{2\pi}\right)
	+ \frac{\sqrt{\pi}}{M L} \mathcal{S}^{(2)} \left(\mu;\frac{M \beta}{4\pi}\right)
	- \frac{\sqrt{\pi}}{M L} \mathcal{S}^{(4)}  (\mu+1; ML)
	+ \frac{\sqrt{\pi}}{M L} \mathcal{S}^{(3)}_0 \left(\mu; \frac{M \beta}{2}\right).
	\label{Eq:Wnu_Full_HalfInteger}
	\end{multline}
	
	Finally, for other real values of the index $\nu$ that are neither integer nor half-integer we have
	\begin{equation}
	{\mathcal W}_\nu^{(1)} = \frac{1}{2} \Bigg[ \frac{\sqrt{\pi}}{M \beta} \Gamma\left(\frac{1}{2}-\nu\right) - \frac{1}{2} \Gamma(-\nu) + \sum_{k=0}^\infty \frac{(-1)^k}{\Gamma(k+1)}\Gamma(\nu-k) \zeta(2\nu-2k) \left(\frac{M\beta}{2}\right)^{-2\nu +2k}
	\Bigg]
	\label{Eq:Wnu_Partial_Real}
	\end{equation}
	\noindent and
	\begin{multline}
	{\mathcal W}_\nu \left(\frac{2}{M^2}\right)^{\nu} = 
	- \frac{1}{4} \Gamma(-\nu) 
	+\frac{\pi}{M^2\beta L}\Gamma\left(1-\nu\right)
	+ \frac{1}{2}\sum_{k=0}^\infty \frac{(-1)^k}{\Gamma(k+1)}\Gamma(\nu-k) \zeta(2\nu-2k) \left[\left(\frac{M\beta}{2}\right)^{-2\nu +2k} + \left(\frac{ML}{2}\right)^{-2\nu +2k}\right]
	\\
	+ \frac{\sqrt{\pi}L}{2\beta} \sum_{k=0}^\infty \frac{(-1)^k}{\Gamma(k+1)}\Gamma\left(\nu-k-\frac{1}{2}\right) \zeta(2\nu-2k-1) \left[\left(\frac{M\beta}{2}\right)^{2k-2\nu} \hspace{-15pt}- \left(\frac{ML}{2}\right)^{2k-2\nu}\right]
	\\+\sum_{k=0}^\infty \frac{(-1)^k}{\Gamma(k+1)} 
	\left(\frac{M \beta}{2\pi}\right)^{2k-2\nu}
	\Bigg[
	\frac{\beta}{2\pi L} \Gamma\left(1-\nu+k\right) \zeta(2-2\nu+2k)
	- \frac{1}{2\sqrt{\pi}} \Gamma\left(\frac{1}{2}-\nu+k\right) \zeta(1-2\nu+2k)
	\Bigg].
	\label{Eq:Wnu_Full_Real}
	\end{multline}
	
	We have managed so far to take into account both situations in which we have one or two compactified dimensions. Let us remember that we are interested in the one-loop Feynman diagram with $\rho$ internal lines and at zero external momentum. The amplitude in the scenario with two compactifications, one related to the inverse temperature $\beta=1/T$ and another to a finite size $L$, is given by Eq.~\eqref{Eq:FuncId2}. And the amplitude in the scenario in which there are $D-1$ dimensions and just one compactification related to the inverse temperature $\beta$ is given by Eq.~\eqref{Eq:FuncId1}
	
	In the following, we investigate, for any value of $\nu$, the relationship between both scenarios. The contribution from $\Gamma(-\nu)$ in amplitude is divergent for integer values of $\nu\ge0$. To avoid the presence of poles in physical quantities we employ the modified minimal subtraction ($\overline{MS}$) scheme~\cite{Weinberg:1996kr} and the function $\Gamma(-\nu)$ is replaced by $\overline{\Gamma}\left(-\nu\right)$ such that
	\begin{equation}
	\overline{\Gamma}\left(-\nu\right) = - \frac{(-1)^{\nu+1}}{\Gamma\left(\nu+1\right)^2} {\nu+1 \brack 2},\quad \nu \ge 0.
	\label{Eq:MMS}
	\end{equation}

	
	\section{Dimensional reduction}
	
	We have considered above a class of one-loop Feynman diagrams with $\rho$ internal lines at a dimension $D$. Their contributions involve the above defined functions $\mathcal{W}_\nu$, where the index $\nu$ is given by $\nu = D/2 - \rho$. In the previous section, we have managed to obtain the final version of $\mathcal{W}_\nu$ in terms of some analytical functions and sums over the Riemann zeta function in the argument. This was done for the specific cases of integer $\nu$ (useful for even dimensions), half-integer values of $\nu$ (for odd dimensions) and for other real values of $\nu$ (for completeness, and which can be considered for models in a fractal dimension).
	
	Taking the situation with $D$ dimensions and letting two of them be compactified, which introduces the temperature $1/\beta$ and a finite length $L$, one might ask how the function behaves as one takes the limit $L\rightarrow0$. This can be interpreted as a ``dimensional reduction". In general, what happens is that the function $\mathcal{W}_\nu (\beta,L)$ diverges as $L\rightarrow 0$. However, if we interpret the procedure of ``dimensional reduction" as taking the dominant contribution\footnote{This is a stronger result for integer dimension $D$ as the divergent terms do not depend on $\beta$. However, this is not the case for a non-integer dimensions $D$ as it can depend on the temperature.} in $\beta$ in the limit $L \rightarrow 0$ and ignore the remaining dependence on the finite length, we obtain the relation
	\begin{equation}
	L \mathcal{I}^{D,2}_\rho \left(M^2;\beta,L\right)\Big|_{L\rightarrow 0} = \mathcal{I}^{D-1,1}_\rho \left(M^2;\beta\right)+ \text{divergent terms}.
	\label{Eq:PrincipalRelation}
	\end{equation}
	\noindent This relation holds for any real value of $D$, which will be shown in the following subsections. The divergent behavior of $\mathcal{I}$ in Eq.~\eqref{Eq:PrincipalRelation} as $L$ goes to zero depends on the quantity $D/2-\rho$.

	
	\subsection{Integer $\nu$, even $D$}
	
	To investigate the so-called dimensional reduction we first consider the case with integer values of $\nu$, which corresponds to even dimensions $D$. The amplitude of a one-loop Feynman diagram in a scenario with both finite temperature and finite size is given by Eq.~\eqref{Eq:FuncId2}. To study its behavior we substitute Eq.~\eqref{Eq:Wnu_Full_Integer} in Eq.~\eqref{Eq:FuncId2} and split its contributions coming from three functions $\mathcal{F}_1,\mathcal{G}_1,\mathcal{H}_1$, such that,
	
	\begin{equation}
	L \mathcal{I}^{D,2}_\rho \left(M^2;\beta,L\right) = \frac{L}{(2\pi)^\frac{D}{2}2^{\rho-2}\Gamma(\rho)} \left[\left(\frac{M^2}{2}\right)^{\frac{D}{2}-\rho} \frac{\Gamma\left(\rho-\frac{D}{2}\right)}{4} + \mathcal{W}_{\frac{D}{2}-\rho}\right]
	= \left(\frac{M^2}{2}\right)^{\frac{D}{2}-\rho} \frac{\mathcal{F}_1 + \mathcal{G}_1 + \mathcal{H}_1}{(2\pi)^\frac{D}{2}2^{\rho-2}\Gamma(\rho)}.
	\end{equation}
	
	The function $\mathcal{H}_1$ is the contribution that vanishes in the $L\rightarrow0$ limit:
	\begin{equation}
	\mathcal{H}_1 = \frac{L}{2} \mathcal{S}^{(2)} \left(\frac{D}{2}-\rho; \frac{ML}{4\pi} \right) + 
	\frac{(-1)^{\frac{D}{2}-\rho}L}{2\Gamma\left(\frac{D}{2}-\rho+1\right)} \left(\gamma+\ln \frac{M L}{4\pi}\right).
	\label{Eq:explicit_H1}
	\end{equation}
	\noindent Note that the zero-temperature contribution was consistently subtracted. The divergent behavior as $L$ goes to zero is given by $\mathcal{G}_1$,
	\begin{equation}
	\mathcal{G}_1 = 
	-\frac{(-1)^{\frac{D}{2}-\rho}}{\Gamma\left(\frac{D}{2}-\rho\right)} \frac{\pi^2}{3M^2L}
	+ \frac{L}{2} \mathcal{S}^{(3)}_1 \left(\frac{D}{2}-\rho; \frac{ML}{2}\right).\label{Eq:explicit_G1}
	\end{equation}
	\noindent Lastly, the function $\mathcal{F}_1$ is the contribution that survives during the \textit{dimensional reduction} and does not diverge,
	\begin{multline}
	\mathcal{F}_1 = - \frac{2\pi}{M^2 \beta} \left[
	\mathcal{S}^{(1)}_1 \left(\frac{D}{2}-\rho; \frac{M\beta}{2\pi}\right)
	+ \mathcal{S}^{(4)} \left(\frac{D}{2}-\rho; M\beta\right)\right]+\frac{\pi \beta}{12} \frac{(-1)^{\frac{D}{2}-\rho}}{\Gamma\left(\frac{D}{2}-\rho+1\right)}\\
	+\begin{cases}
	\frac{\pi}{M^2 \beta} \Gamma\left(1-\frac{D}{2}+\rho\right) &, \frac{D}{2} \le \rho\\
	\frac{\pi}{M^2 \beta} \frac{(-1)^{\frac{D}{2}-\rho}}{\Gamma\left(\frac{D}{2}-\rho\right)} \left(-\frac{1}{\Gamma\left(\frac{D}{2}-\rho\right)}{\frac{D}{2}-\rho \brack 2}  
	+ 2 \ln M\beta\right)
	&, \frac{D}{2}> \rho
	\end{cases}.\label{Eq:explicit_F1}
	\end{multline}
	
	Therefore, as the $L\rightarrow 0$ limit is taken the contribution $\mathcal{H}_1$ vanishes and the divergent behavior in $L$ is given by the function $\mathcal{G}_1$:
	
	\begin{equation}
	L \mathcal{I}^{D,2}_\rho \left(M^2;\beta,L\right) \Big|_{L\rightarrow 0} = \frac{1}{(2\pi)^\frac{D}{2}2^{\rho-2}\Gamma(\rho)}L \mathcal{W}_{\frac{D}{2}-\rho} \Big|_{L\rightarrow 0} = \left(\frac{M^2}{2}\right)^{\frac{D}{2}-\rho} \frac{\mathcal{F}_1 + \mathcal{G}_1}{(2\pi)^\frac{D}{2}2^{\rho-2}\Gamma(\rho)}.
	\end{equation}
	
	On the other hand, let us consider the second case in which we start from a scenario with one less dimension, $D-1$. The amplitude of the one-loop Feynman diagram is simply Eq.~\eqref{Eq:FuncId1}, here written as
	\begin{equation}
	\mathcal{I}^{D-1,1}_\rho(M^2; \beta)
	= \frac{\left(\frac{M^2}{2}\right)^{\frac{D}{2}-\rho}}{(2\pi)^{\frac{D}{2}} 2^{\rho-2}\Gamma[\rho]} \left\{
	\frac{\sqrt{\pi}}{2M}\Gamma\left(\rho-\frac{D}{2}+\frac{1}{2}\right)
	+ \frac{2\sqrt{\pi}}{M} {\mathcal W}^{(1)}_{\frac{D}{2}-\frac{1}{2}-\rho} \left(M^2; \beta\right)
	\right\}.
	\label{Eq:EvenIopen}
	\end{equation}
	\noindent And, for even dimensions $D$, we get that $\frac{D}{2}-\frac{1}{2}-\rho$ is a half-integer. So, we use Eq.~\eqref{Eq:Wnu_Partial_HalfInteger} with $\mu = \frac{D}{2}-\rho-1$ and obtain
	
	\begin{multline}
	{\mathcal W}_{\frac{D}{2}-\frac{1}{2}-\rho}^{(1)} = -\frac{1}{4} \Gamma\left(\frac{1}{2}-\frac{D}{2}+\rho\right) 
	+ \frac{\sqrt{\pi}}{2M \beta} \Gamma\left(-\frac{D}{2}+\rho+1\right)
	+ \frac{\sqrt{\pi}}{2M\beta} \frac{(-1)^{\frac{D}{2}-\rho-1}}{\Gamma\left(\frac{D}{2}-\rho\right)} \left( {\frac{D}{2}-\rho \brack 2} \frac{1}{\Gamma\left(\frac{D}{2}-\rho\right)} - 2\ln M \beta \right) \\
	- \frac{\sqrt{\pi}}{M \beta} \mathcal{S}^{(4)} \left(\frac{D}{2}-\rho;M\beta\right) 
	- \frac{\sqrt{\pi}}{M \beta} \mathcal{S}^{(1)}_0 \left(\frac{D}{2}-\rho;\frac{M \beta}{2\pi}\right).
	\end{multline}
	\noindent Therefore, Eq.~\eqref{Eq:EvenIopen} becomes
	\begin{multline}
	\mathcal{I}^{D-1,1}_\rho(M^2; \beta)
	= \frac{\left(\frac{M^2}{2}\right)^{\frac{D}{2}-\rho}}{(2\pi)^{\frac{D}{2}} 2^{\rho-2}\Gamma[\rho]} \left\{
	\frac{\pi}{M^2 \beta} \Gamma\left(\rho-\frac{D}{2}+1\right) 
	+ \frac{\pi}{M^2 \beta} \frac{(-1)^{\frac{D}{2}-\rho-1}}{\Gamma\left(\frac{D}{2}-\rho\right)} \left( {\frac{D}{2}-\rho \brack 2} \frac{1}{\Gamma\left(\frac{D}{2}-\rho\right)} - 2\ln M \beta \right) \right.\\\left.
	- \frac{2\pi}{M^2 \beta} \mathcal{S}^{(4)} \left(\frac{D}{2}-\rho;M\beta\right) 
	- \frac{2\pi}{M^2 \beta} \mathcal{S}^{(1)}_0 \left(\frac{D}{2}-\rho;\frac{M \beta}{2\pi}\right)
	\right\}.
	\end{multline}
	
	Furthermore, since $\frac{2\pi}{M^2 \beta} \mathcal{S}^{(1)}_0 \left(\frac{D}{2}-\rho;\frac{M \beta}{2\pi}\right) = \frac{2\pi}{M^2 \beta} \mathcal{S}^{(1)}_1 \left(\frac{D}{2}-\rho;\frac{M \beta}{2\pi}\right) - \frac{\pi\beta}{12} \frac{(-1)^{\frac{D}{2}-\rho}}{\Gamma\left(\frac{D}{2}-\rho+1\right)}$, by direct comparison with $\mathcal{F}_1$ we get
	\begin{equation}
	\mathcal{I}^{D-1,1}_\rho(M^2; \beta)
	= \frac{\left(\frac{M^2}{2}\right)^{\frac{D}{2}-\rho}}{(2\pi)^{\frac{D}{2}} 2^{\rho-2}\Gamma[\rho]} 
	\mathcal{F}_1.
	\end{equation}
	\noindent Then we find the relation
	\begin{equation}
	L \mathcal{I}^{D,2}_\rho \left(M^2;\beta,L\right) \Big|_{L\rightarrow 0} = \mathcal{I}^{D-1,1}_\rho \left(M^2;\beta\right) +\left(\frac{M^2}{2}\right)^{\frac{D}{2}-\rho} \frac{\mathcal{G}_1}{(2\pi)^\frac{D}{2}2^{\rho-2}\Gamma(\rho)}
	= \mathcal{I}^{D-1,1}_\rho \left(M^2;\beta\right) + \mathcal{O}\left(L^{-1}\right)
	\end{equation}
	
	The asymptotic behavior of the function $\mathcal{G}_1$ as $L\rightarrow 0$ is 
\begin{align*}
\mathcal{G}_1 \sim 
\begin{cases}
L^{-1} &, D-2\rho < 4; 
\\
L^{-(D-2\rho-1)} &, D-2\rho \ge 4.
\end{cases}
\end{align*}

	
	\subsection{Half-integer $\nu$, odd $D$}
	
	Now we consider odd dimensions $D$, meaning half-integer values of $\nu = \mu+1/2$. For reference, $\mu = \frac{D-1}{2}-\rho$.  Again, we split the contributions in three functions, $\mathcal{F}_2, \mathcal{G}_2, \mathcal{H}_2$, and after substitution of Eq.~\eqref{Eq:Wnu_Full_HalfInteger} in Eq.~\eqref{Eq:FuncId2} the amplitude is given by
	\begin{equation}
	L \mathcal{I}^{D,2}_\rho \left(M^2;\beta,L\right)= \left(\frac{M^2}{2}\right)^{\frac{D}{2}-\rho} \frac{1}{(2\pi)^\frac{D}{2}2^{\rho-2}\Gamma(\rho)} \left(\mathcal{F}_2 + \mathcal{G}_2 + \mathcal{H}_2\right),
	\end{equation}
	\noindent with
	\begin{align}
	\mathcal{F}_2 =&
	\frac{\pi}{M^2 \beta} \Gamma \left(1-\frac{D}{2}+\rho\right)
	+ \frac{\sqrt{\pi}}{M} \frac{(-1)^{\frac{D-1}{2}-\rho}}{\Gamma\left(\frac{D+1}{2}-\rho\right)} \left(\gamma+\ln \frac{M \beta}{4\pi}\right) \nonumber
	\\& + \frac{\sqrt{\pi}}{M} \mathcal{S}^{(2)} \left(\frac{D-1}{2}-\rho; \frac{M \beta}{4\pi}\right)
	+ \frac{\sqrt{\pi}}{M} \mathcal{S}^{(3)}_0 \left(\frac{D-1}{2}-\rho;\frac{M \beta}{2}\right)\label{Eq:explicit_F2},
	\\
	\mathcal{G}_2 =& 
	- \frac{\sqrt{\pi}}{M}\frac{(-1)^{\frac{D-1}{2}-\rho}}{\Gamma\left(\frac{D+1}{2}-\rho\right)} \ln ML
	- \frac{\sqrt{\pi}}{M} \mathcal{S}^{(4)}\left(\frac{D+1}{2}-\rho;ML\right) \label{Eq:explicit_G2},
	\\
	\mathcal{H}_2 =& 
	-\frac{\sqrt{\pi}}{M} \mathcal{S}^{(1)}_0\left(\frac{D+1}{2}-\rho;\frac{M L}{2\pi}\right) \label{Eq:explicit_H2}.
	\end{align}
	
	The second scenario, with a reduced number of dimensions, is given by Eq.~\eqref{Eq:FuncId1}:
	\begin{equation}
	\mathcal{I}^{D-1,1}_\rho(M^2; \beta)
	= \frac{\left(\frac{M^2}{2}\right)^{\frac{D}{2}-\rho}}{(2\pi)^{\frac{D}{2}} 2^{\rho-2}\Gamma[\rho]} \left\{
	\frac{\sqrt{\pi}}{2M}\Gamma\left(\rho+\frac{1-D}{2}\right)
	+ \frac{2\sqrt{\pi}}{M} {\mathcal W}^{(1)}_{\frac{D-1}{2}-\rho} \left(M^2; \beta\right)
	\right\},
	\end{equation}
	\noindent with ${\mathcal W}^{(1)}$ given by Eq.~\eqref{Eq:Wnu_Partial_Integer}, that is,
	\begin{multline}
	{\mathcal W}_{\frac{D-1}{2}-\rho}^{(1)} = 
	- \frac{1}{4} \Gamma\left(\rho+\frac{1-D}{2}\right)
	- \frac{(-1)^{\frac{D-1}{2}-\rho}}{4\Gamma\left(\frac{D+1}{2}-\rho\right)^2} {\frac{D+1}{2}-\rho \brack 2}
	+\frac{\sqrt{\pi}}{2M\beta} \Gamma\left(1+\rho-\frac{D}{2}\right) 
	+ \frac{(-1)^{\frac{D-1}{2}-\rho}}{2\Gamma\left(\frac{D+1}{2}-\rho\right)} \left(\gamma+\ln \frac{M \beta}{4\pi}\right)\\
	+ \frac{1}{2} \mathcal{S}^{(3)}_0 \left(\frac{D-1}{2}-\rho;\frac{M\beta}{2}\right)
	+ \frac{1}{2} \mathcal{S}^{(2)}\left(\frac{D-1}{2}-\rho;\frac{M\beta}{4\pi}\right).
	\end{multline}
	
	Once again, we must be careful with the zero-temperature contribution, as a modified minimal subtraction scheme is assumed, see Eq.~\eqref{Eq:MMS}. We obtain, after the cancellation with the zero-temperature contribution, 
	\begin{equation}
	\mathcal{I}^{D-1,1}_\rho(M^2; \beta)
	= \frac{\left(\frac{M^2}{2}\right)^{\frac{D}{2}-\rho}}{(2\pi)^{\frac{D}{2}} 2^{\rho-2}\Gamma[\rho]}  \mathcal{F}_2.
	\end{equation}
	
	Therefore, we get the relation
	\begin{equation}
	L \mathcal{I}^{D,2}_\rho \left(M^2;\beta,L\right) \Big|_{L\rightarrow 0} = \mathcal{I}^{D-1,1}_\rho 
	\left(M^2;\beta\right) +\left(\frac{M^2}{2}\right)^{\frac{D}{2}-\rho} \frac{\mathcal{G}_2}{(2\pi)^\frac{D}{2}2^{\rho-2}\Gamma(\rho)}
	= \mathcal{I}^{D-1,1}_\rho \left(M^2;\beta\right)
	+ \mathcal{O}\left(\ln L\right).
	\end{equation}
	
	The only difference compared to the scenario of integer values of $\nu$ is the divergent behavior of $\mathcal{G}_2$, in this case we have
	\begin{align*}
\mathcal{G}_2 \sim 
\begin{cases}
\ln L &, D-2\rho < 3; 
\\
L^{-(D-2\rho-1)} &, D-2\rho \ge 3.
\end{cases}
\end{align*}
	
	
	\subsection{Other real $\nu$, non-integer $D$}
	
	To complete the analysis, we consider other real values of $\nu$, that allow to take into account non-integer values of the dimension $D$. We follow the same procedure, defining three functions, $\mathcal{F}_3, \mathcal{G}_3, \mathcal{H}_3$; after substituting Eq.~\eqref{Eq:Wnu_Full_Real} into Eq.~\eqref{Eq:FuncId2}, we get
	\begin{equation}
	L \mathcal{I}^{D,2}_\rho \left(M^2;\beta,L\right) = \left(\frac{M^2}{2}\right)^{\frac{D}{2}-\rho} \frac{1}{(2\pi)^\frac{D}{2}2^{\rho-2}\Gamma(\rho)} \left(\mathcal{F}_3 + \mathcal{G}_3 + \mathcal{H}_3\right),
	\end{equation}
	\begin{align}
	\mathcal{F}_3 &= 
	\frac{\pi}{M^2\beta}\Gamma\left(1-\nu\right)
	+ \sum_{k=0}^\infty \frac{(-1)^k}{\Gamma(k+1)}     \left(\frac{M \beta}{2\pi}\right)^{2k-2\nu}    \frac{\beta}{2\pi} \Gamma\left(1-\nu+k\right) \zeta(2-2\nu+2k) \label{Eq:explicit_F3},
	\\
	\mathcal{G}_3 &= 
	\frac{L}{2}\sum_{k=0}^{\floor{\frac{\nu-1}{2}}} \frac{(-1)^k}{\Gamma(k+1)}\Gamma(\nu-k) \zeta(2\nu-2k) \left(\frac{ML}{2}\right)^{-2\nu +2k} \nonumber
	\\&- \frac{\sqrt{\pi}L^2}{2\beta} \sum_{k=0}^{\floor{\nu-1}} \frac{(-1)^k}{\Gamma(k+1)}\Gamma\left(\nu-k-\frac{1}{2}\right) \zeta(2\nu-2k-1) \left(\frac{ML}{2}\right)^{2k-2\nu}\label{Eq:explicit_G3},
	\\
	\mathcal{H}_3 &= 
	\frac{L}{2}\sum_{k=0}^\infty \frac{(-1)^k}{\Gamma(k+1)}\Gamma(\nu-k) \zeta(2\nu-2k) \left(\frac{M\beta}{2}\right)^{-2\nu +2k} \nonumber \\&
	+ \frac{L}{2}\sum_{k=\max\left(0,\floor{\frac{\nu-1}{2}}\right)}^\infty \frac{(-1)^k}{\Gamma(k+1)}\Gamma(\nu-k) \zeta(2\nu-2k) \left(\frac{ML}{2}\right)^{-2\nu +2k}\nonumber\\&
	+ \frac{\sqrt{\pi}L^2}{2\beta} \sum_{k=0}^\infty \frac{(-1)^k}{\Gamma(k+1)}\Gamma\left(\nu-k-\frac{1}{2}\right) \zeta(2\nu-2k-1) \left(\frac{M\beta}{2}\right)^{2k-2\nu}\nonumber\\&
	- \frac{\sqrt{\pi}L^2}{2\beta} \sum_{k=\max\left(0,\floor{\nu-1}\right)}^\infty \frac{(-1)^k}{\Gamma(k+1)}\Gamma\left(\nu-k-\frac{1}{2}\right) \zeta(2\nu-2k-1) \left(\frac{ML}{2}\right)^{2k-2\nu}\nonumber\\&
	-\frac{L}{2\sqrt{\pi}} \sum_{k=0}^\infty \frac{(-1)^k}{\Gamma(k+1)} 
	\left(\frac{M \beta}{2\pi}\right)^{2k-2\nu}
	\Gamma\left(\frac{1}{2}-\nu+k\right) \zeta(1-2\nu+2k).\label{Eq:explicit_H3}
	\end{align}
	
    The second scenario is, in this case, obtained after substitution of Eq.~\eqref{Eq:Wnu_Partial_Real} into Eq.~\eqref{Eq:FuncId1},
	\begin{equation*}
	\mathcal{I}^{D-1,1}_\rho \left(M^2;\beta\right) = 
	\frac{\left(\frac{M^2}{2}\right)^{\frac{D}{2}-\rho}}{(2\pi)^\frac{D}{2}2^{\rho-2}\Gamma(\rho)}  \Bigg[ 
	\frac{\pi}{M^2 \beta} \Gamma\left(1-\nu\right)
	+ \frac{\sqrt{\pi}}{M}\sum_{k=0}^\infty \frac{(-1)^k}{\Gamma(k+1)}\Gamma\left(\nu-k-\frac{1}{2}\right) \zeta(2\nu-2k-1) \left(\frac{M\beta}{2}\right)^{-2\nu +2k+1}
	\Bigg],
	\end{equation*}
	\noindent The product of the gamma and zeta functions can be rewritten and we obtain
	\begin{equation}
	\mathcal{I}^{D-1,1}_\rho \left(M^2;\beta\right) = \left(\frac{M^2}{2}\right)^{\frac{D}{2}-\rho} \frac{1}{(2\pi)^\frac{D}{2}2^{\rho-2}\Gamma(\rho)} \mathcal{F}_3.
	\end{equation}
	
	Finally, we obtain the simple relation
	
	\begin{equation}
	L \mathcal{I}^{D,2}_\rho \left(M^2;\beta,L\right)\Big|_{L\rightarrow 0} = \mathcal{I}^{D-1,1}_\rho 
	\left(M^2;\beta\right) +\left(\frac{M^2}{2}\right)^{\frac{D}{2}-\rho} \frac{\mathcal{G}_3}{(2\pi)^\frac{D}{2}2^{\rho-2}\Gamma(\rho)}
	= \mathcal{I}^{D-1,1}_\rho\left(M^2;\beta\right)
	+ \mathcal{O}\left(L\right).
	\end{equation}
	
	The leading behavior on $L$ depends on the structure of $\mathcal{G}_3$ which has the asymptotic behavior $\mathcal{G}_3 \sim L^{-(D-2\rho-1)}, D-2\rho \ge 2$ as $L\rightarrow $.

	\section{Conclusions}
	
	We obtained that, at least for the class of one-loop diagrams, it is possible to consistently reduce the dimension of the system if one proceeds carefully.
	
	We are not allowed to perform a strict dimensional reduction taking a model with a finite length $L$ and suppressing it continuously to zero. The first obstruction is the function $\mathcal{G}$, that carries the divergent behavior as $L$ goes to zero. Of course, strictly speaking, $\mathcal{I}^{D,2}_\rho\left(M^2;\beta,L\right)$ cannot be evaluated at $L=0$ due to this divergence. However, we can also obtain some specific possibilities where $\mathcal{G}=0$ which occurs for $D=1,2$ or any non-integer dimension $D<4$. Of course, $D=1$ is inconsistent (as there are 2 compactified dimensions) and must be discarded. And $D=2$ is the case where all dimensions are compactified. Anyway, assuming $\mathcal{G}=0$ the procedure of dimensional reduction for $L\rightarrow 0$ is
	\begin{equation}
	L \mathcal{I}^{D,2}_\rho \left(M^2;\beta,L\right)\Big|_{L\rightarrow 0} = \mathcal{I}^{D-1,1}_\rho 
	\left(M^2;\beta\right). \quad \left(\text{for} \quad \frac{D}{2}-\rho<1\right)
	\label{Eq:RelationG0}
	\end{equation}
	\noindent The identification in Eq.~\eqref{Eq:RelationG0} is also valid for any $D$ if we choose the prescription to ignore the divergent function $\mathcal{G}$. We define that this prescription is what we call a ``dimensional reduction": we take the limit of a function as the length $L$ goes to zero and remove its divergent components.
	
	We remark that the $L$ on the right side of the above equation is simply indicating that $\mathcal{I}^{D,2}_\rho$ also diverges as $L$ goes to zero despite the existence of $\mathcal{G}$. The result that a continuous approach to the dimensional reduction is not possible is not a surprise. Indeed, in previous articles, in the context of phase transitions, it has been found the presence of a \textit{minimal length} below which the phase transition ceases to occur. Moreover, this result agrees with experimental observations about the existence of a minimal length. 
	
	Notwithstanding, now that we made clear that a strictly dimensional reduction is not attainable, we are allowed to discuss the existence of a prescription to do so. The idea is that there is a relationship between both situations: one with a small system length $L$ and another where this dimension is ignored from the beginning.
	
	Moreover, had we considered a $N$ component scalar model with quartic interaction in $D=3$ with tree-level coupling $\lambda$ and mass $m$. This describe a heated surface. Taking the large $N$ limit and using a formal resumation, the first loop correction to the coupling constant $g$ and squared mass $M^2$ are
    \begin{subequations}
	\begin{align}
	g_3 = \frac{\lambda_3}{1 - \lambda_3 \mathcal{I}^{3,1}_2\left(M^2;\beta\right)},\\
	M^2 = m^2 + \lambda_3 \mathcal{I}^{3,1}_1\left(M^2;\beta\right).
    \end{align}
	\label{Eq:ExampleD3}
    \end{subequations}
    \noindent Assuming that, indeed, the surface is a ``dimensionally reduced" case of a heated film with thickness $L$ we can ignore the divergent component $\mathcal{G}$ and use the identification in Eq.~\eqref{Eq:RelationG0} to write Eqs.~\eqref{Eq:ExampleD3} as
    \begin{subequations}
	\begin{align}
	g_3 = \frac{\lambda_3}{1 - (\lambda_3 L) \mathcal{I}^{4,2}_2\left(M^2;\beta,L\right)},\\
	M^2 = m^2 + (\lambda_3 L) \mathcal{I}^{4,2}_1\left(M^2;\beta,L\right).
	\end{align}
    \label{Eq:ExampleD3mod}
    \end{subequations}
	\noindent On the other hand, if we simply consider a heated film in $D=4$ with $N$ scalar fields and explore its behavior for very small thickness we get
    \begin{subequations}
	\begin{align}
	g_4 = \frac{\lambda_4}{1 - \lambda_4 \mathcal{I}^{4,2}_2\left(M^2;\beta,L\right)},\\
	M^2 = m^2 + \lambda_4 \mathcal{I}^{4,2}_1\left(M^2;\beta,L\right).
	\end{align}
	\label{Eq:ExampleD4}
    \end{subequations}
	Comparing both scenarios in Eqs.~\eqref{Eq:ExampleD3mod} and Eqs.~\eqref{Eq:ExampleD4}, we see that the coupling constant from the planar scenario (both the free $\lambda_3$ and corrected $g_3$) is related with the coupling constant from the thin film scenario by
	\begin{equation}
	\lambda_3 L = \lambda_4.
	\end{equation}
	\noindent We can extract from this simple relation some important conclusions:
	\begin{itemize}
		\item The strict dimensional reduction, once again, is not allowed. It would require that the coupling constant for the reduced scenario goes to infinity.
		\item This is a direct indication that both scenarios are physically different.
		\item In the context of phase transitions or some other situation where it can be observed the existence of a minimal thickness $L_{\text{min}}$ we can consider that the effective coupling constant in the planar scenario is $\lambda_3 = \lambda_4/L_{\text{min}}$.
	\end{itemize}

	\bibliography{Q_Refs}
\end{document}